# Magnetic wire as stress controlled micro-rheometer for cytoplasm viscosity measurements

Jean-François Berret
Matière et Systèmes Complexes, UMR 7057 CNRS Université Denis Diderot Paris-VII, Bâtiment Condorcet, 10 rue Alice Domon et Léonie Duquet, F-75205 Paris, France

**ABSTRACT**

We review here different methods to measure the bulk viscosity of complex fluids using micron-sized magnetic wires. The wires are characterized by length of a few microns and diameter of a few hundreds of nanometers. We first draw analogies between cone-and-plate rheometry and wire-based microrheology. In particular we highlight that magnetic wires can be operated as stress-controlled rheometers for two types of testing, the creep-recovery and steady shear experiments. In the context of biophysical applications, the cytoplasm of different cell lines including fibroblasts, epithelial and tumor cells is studied. It reveals that the interior of living cells can be described as a viscoelastic liquid with a static viscosity comprised between 10 and 100 Pas. We extend the previous approaches and show that the proposed technique can also provide time resolved viscosity data, which for cells display strong temporal fluctuations. The present work demonstrates the high potential of the magnetic wires for quantitative rheometry in confined espaces.

## 1. INTRODUCTION

In 2008 our group reported for the first time the fabrication of rigid and anisotropic magnetic wires obtained from bottom-up assembly using iron oxide nanoparticles.[1] The first syntheses resulted in objects with length of a few microns and diameter of a few hundreds of nanometers. Since then, the preparation method was further improved and wires up to 500 µm in length and 2 µm in diameter, functionalized either at the surface or in the core have been made.[2,3] The building blocks used for the fabrication are polyelectrolytes, such as poly(diallyldimethylammonium chloride) or poly(ethylenimine) and 10 nm maghemite ($\gamma$-$Fe_2O_3$) nanocrystals, both species being available commercially. The co-assembly process is based on the dialysis of water borne dispersions with an excess of salt. For the linear growth of the aggregates, the dialysis bath is placed in a permanent magnetic field of around 100 mT generated by rare-earth or magnetite permanent magnets. This method is versatile and provides large amounts of objects, around $10^{10}$ per dialysis batch. In terms of structure, small-angle X-ray scattering has shown that the particles are rigidly attached to each other and in close contact, resulting in high volume fractions of magnetic material, between 30% and 40%.[3] These findings were later corroborated by transmission electron microscopy imaging.[2] Following these first steps, we developed applications for the wires as sensors and microactuators.[4-6] In the context of microrheology,[7-12] it appears that the magnetic wires could be of high interest and provide tools for quantitative viscoelasticity measurements.[13-20]

When put in a constant magnetic field, the wires are submitted to a magnetic torque $\Gamma_M$ of the form:[21,22]

$$\Gamma_M(\theta, H) = \frac{1}{2}\mu_0 V \Delta\chi H^2 \sin(2\theta) \qquad (1)$$

where $\mu_0$ the permeability in vacuum, $V$ is the volume of the wire ($V = \frac{\pi}{4}D^2 L$, $L$ length, $D$ diameter), $\Delta\chi$ the anisotropy of susceptibility between parallel and perpendicular directions, $H$ the magnetic excitation ($H = B/\mu_0$ where B is the magnetic field) and $\theta$ is the angle between the wire and the field orientation. Under the application of a magnetic torque $\Gamma_M$, the wire rotates in a propeller-like motion so as to minimize $\theta$ and to eventually align with $H$.

Eq. 1 was first derived in 1990 by Helgesen and coworkers tracking the motion of bound pairs of non-magnetic microspheres immersed in a ferrofluid medium.[21,22] It was shown that the bound microsphere model was also appropriate to describe the wire mechanical behavior.[4,23] For an overview of complex fluids investigated with this technique, we refer to recent reviews in Refs.[9,11,12] An important feature in Eq. 1 is the quadratic dependence of the torque $\Gamma_M \sim H^2$, meaning that the wires have kept of the superparamagnetic character of the nanoparticles (for ferromagnetic wires[13] the torque varies linearly with H). This property has a crucial incidence: knowing the amplitude and direction of the field, and knowing the wire orientation, $\Gamma_M$ can be evaluated. It is hence possible to calculate the forces exerted to the surrounding fluid and to develop a quantitative microrheology approach. The only parameter to be determined beforehand is the susceptibility anisotropy $\Delta\chi$ characteristic of the magnetic material.[4,21-23] Here we review different methods to measure the viscosity of soft matter using magnetic wire manipulation. Some representative data obtained on complex fluids including water-glycerol mixtures, viscoelastic surfactant solutions or the intracellular medium of living cells are described. We also discuss the possibility to exploit these wires as probes for passive micro-rheology.

## 2. PASSIVE MICRO-RHEOLOGY

With passive micro-rheology, no magnetic field is applied and the Brownian motion of the wires is monitored by optical microscopy using a high-speed camera. Here we focus on the rotational Brownian motion of the objects and do not discuss their translational diffusion. The challenge is here to get the three-dimensional trajectories of anisotropic probes with optical techniques that essentially provide two-dimensional information. To solve this problem, the apparent length and the angular displacement of the wire projected in the focal plane of the objective are tracked as a function of the time, allowing to determine the wire orientation spherical coordinates. This procedure is achieved using a homemade tracking algorithm implemented as an ImageJ plugin (https://imagej.nih.gov/ij/). From these coordinates, a new angular variable $\psi(t)$ is defined. It was shown theoretically that $\psi(t)$ obey the Langevin diffusion equation and that the mean-squared angular displacement (MSAD) associated to the $\psi$–variable increases linearly with the lag time, as:

$$\langle \Delta\psi^2(t,L) \rangle = 2D_{rot}(L)t$$

$$\text{where} \quad D_{rot}(L) = \frac{3k_BT}{\pi L^3 \eta} g\left(\frac{L}{d}\right) \tag{2}$$

Here, $k_BT$ the thermal energy, $k_B$ Boltzmann constant and $g(p) = \ln(p) - 0.662 + 0.917\,p - 0.050\,p^2$.[4] The surfactant solution studied here is a cetylpyridinium chloride/sodium salicylate (CPCl/NaSal) dispersion, which is known to display a Maxwell behavior in rheology.[24,25] Fig. 1a shows a 6.6 µm long wire immersed in a CPCl/NaSal wormlike micellar solution at the concentration $c = 1$ wt. %.

The wire orientation is determined by the angle $\varphi$ which time dependence is illustrated in Fig. 1b over a period of 20 s. The 3D Brownian motion is extracted from its 2D projection according the procedure described previously.[5] Fig. 1c shows the MSADs of wires of different lengths ($L = 5, 12, 21$ and $25$ µm) immersed in the surfactant solution as a function of the lag time. At short times, the MSADs exhibit a slow increase followed by a linear dependence, indicated by the straight lines.

The data also emphasizes that the slope in the linear regime is strongly varying with $L$. The $\langle \Delta\psi^2(t) \rangle$-data were adjusted with Eq. 2 for 20 different wires, each experiment being performed in duplicate. From this data set, we could show that the rotational diffusion constant varies as $g\left(\frac{L}{d}\right)L^{-3}$. At 20 °C, the static shear viscosity was found to be $\eta = 0.20 \pm 0.08$ Pa s, in excellent agreement with the value determined by cone-and-plate rheometry ($0.14 \pm 0.05$ Pa s).[25] In conclusion to this part, we have shown that tracking the Brownian rotational motion of wires provides a new and reliable method to study the micro-rheology of simple and complex fluids. Due to their large aspect ratio, these objects can probe complex fluids over different length scales and be sensitive to heterogeneities. The technique allows the measure of the static viscosity from 1 to $10^6$ times the water viscosity.[5,15,20]

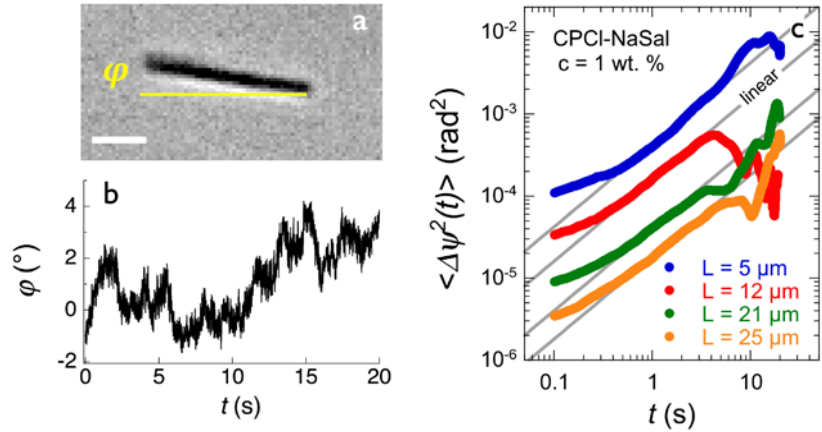

**Figure 1.** a) Image of a 6.6 μm long wire immersed in a CPCl/NaSal wormlike micellar solution at the concentration $c$ = 1 wt. % and temperature 20 °C. b) Time dependence of the angle $\varphi(t)$ in a typical experiment. The time interval between two consecutive point is 1 ms. c) Mean-squared angular displacement *versus* lag time for wires of different lengths. The straight lines in gray indicate a linear variation.

## 3. ACTIVE MICRO-RHEOLOGY

Typical rheology testing using a stress controlled rheometer coupled with a cone-and-plate or Couette device are *i)* creep-recovery experiment, where a stress is applied stepwise to the sample and the resulting deformation is monitored as a function of the time, *ii)* dynamic frequency sweep where the imposed stress oscillates at increasing frequencies, and *iii)* steady shear experiment where a constant shear stress is applied to the solution until the steady state is reached. The same tests can be achieved with magnetic wires. In this paper we will review the experimental configurations corresponding to the creep-recovery and steady shear experiments, that is the cases *i)* and *iii)* above.

### 3.1 Transient creep-recovery experiments

In creep-recovery experiments, a constant shear stress is applied to a sample for a given time interval and the deformation rate is measured *versus* time. With micron-sized wires, the magnetic field is switched between two $\pi/2$ orientations, as indicated in Fig. 2a. In the experiment, the wire is first oriented by the pair of coils along the Y-direction. Once aligned, this primary field is switched off at time $t_0$ whereas the field along the X-direction is switched on. As a result, the wire rotates in a propeller-like motion and orients along OX. In this experiment the principle is similar to the creep-recovery test of rheology, except that the torque (given by Eq. 1) is changing continuously over time as the wire reorients. The monitoring of the angle during its rotation shows a linear decrease at the field inception, followed by a progressive alignment (Fig. 2b). The torque balance equation leads to a time-dependence of the form: $tg(\theta(t)) = tg(\theta_0)exp(-k(t - t_0))$, where the decay rate is given by:[1]

$$k = \frac{\mu_0 \Delta \chi}{2\eta} g\left(\frac{L}{D}\right) \frac{D^2}{L^2} H^2 \qquad (3)$$

The data in Fig. 2b were obtained for a 14.3 μm wire moving in deionized water under a magnetic field of 3.1 mT. The continuous line in the figure shows a good agreement between the data and the model. The transient creep-recovery experiments were performed on different wires at various magnetic fields from 2 to 10 mT. The decay rate $k$ retrieved from the fitting is shown *versus* $H^* = \sqrt{g\left(\frac{L}{D}\right)} DH/L$ in Fig. 2c in double logarithmic scale to emphasize a possible power law variation. The data are indeed following a quadratic dependence of the form $k = \frac{\mu_0 \Delta \chi}{2\eta} H^{*2}$ predicted by the model. From the slope, the anisotropy of susceptibility $\Delta\chi$ can be found. In this particular case, we obtained $\Delta\chi = 0.40 \pm 0.02$, in fair agreement with previous determinations.[1,6]

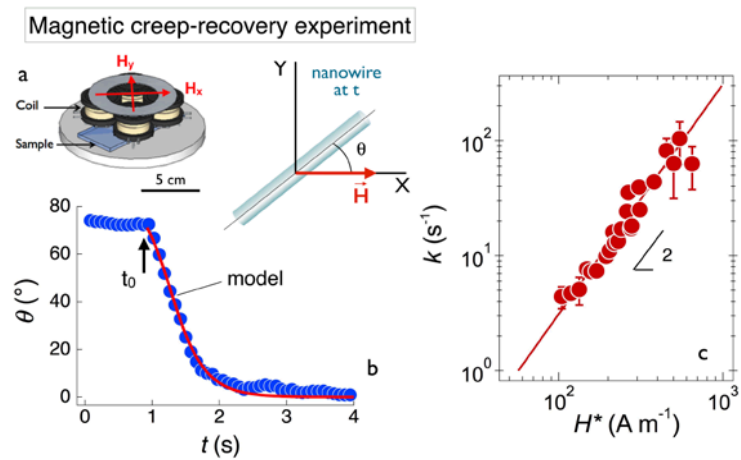

**Figure 2.** a) Left panel: side view of the rotating field device used to actuate the magnetic wires. Right panel: schematic representation of the wire reorienting along the X-axis. b) Orientation angle $\theta(t)$ of a 14.3 μm wire in deionized water at room temperature. The continuous line corresponds to a model explained in the text. c) Decay rate $k$ as a function of the reduced parameter $H^* = H\sqrt{g(p)}/p$ where $p = L/D$ denotes the wire anisotropy ratio. The susceptibility $\Delta\chi$ (= 0.40 ± 0.02) is determined from the $k(H^*)$ quadratic dependence assuming a value for the water viscosity of $0.89\times10^{-3}$ Pa s.

### 3.2 Steady rotation experiment

In rheometry, steady stress experiments correspond to conditions where a constant shear stress is applied to the sample and results in a steady deformation rate. From the values of the imposed shear stress and those of the measured shear rates, the material flow curve can be constructed. To account for the specificity of the wire magnetic properties, a steady rotation can be achieved by applying a rotating magnetic field, as first suggested by Helgesen and coworkers.[21,22] The magnetic wire-based microrheology technique using rotating field is termed Magnetic Rotational Spectroscopy (MRS) and has been exploited for microrheology studies of various complex fluids.[9-12]

Fig. 3a illustrates the wire rotation in water. The different images of this chronophotograph are taken at fixed time interval during a 180° spin, showing that magnetic wires do rotate with the field. In simple Newton fluid however, a wire also experiences a restoring torque that slows down its motion. With increasing frequency, this friction torque increases and above a critical value $\omega_C$ the wire cannot follow the field any more. The wire then undergoes a transition between a synchronous and an asynchronous regime at:[21-23]

$$\omega_C = \frac{3}{8}\frac{\mu_0 \Delta\chi}{\eta_0} g\left(\frac{L}{D}\right)\frac{D^2}{L^2}H^2 \qquad (4)$$

The previous feature is specific to magnetic wires, but not to cone-and-plate stress controlled rheometer. The critical frequency $\omega_C$ in Eq. 4 relates to the decay rate found in the creep-recovery measurements through $\omega_C = 4k/3$.

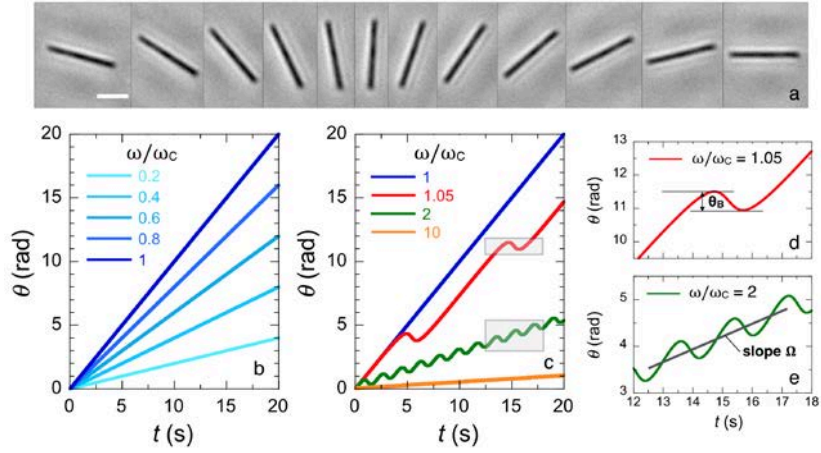

**Figure 3.** a) Chronophotograph of a 10 μm wire undergoing a 180° rotation. The bar is 5 μm. b) Time dependence of the rotation angle $\theta(t)$ at frequencies below the critical frequency $\omega_C$ in the synchronous regime. c) Same as in Fig. 3b for $\omega \geq \omega_C$. In the asynchronous regime, the wire exhibits back-and-forth oscillations. The rectangles in gray refer to Fig. 3d and 3e, respectively. d) Expanded view of an oscillation at $\omega/\omega_C = 1.05$ highlighting the angle $\theta_B$. e) Expanded view of several oscillation at $\omega/\omega_C = 2$ showing the average angular velocity $\Omega$.

Fig. 3b and 3c display the generic behaviors in the two regimes derived from the mechanical constitutive equation. At low frequency (Fig. 3b), the wire rotates in phase with the field and $\theta(t) = \omega t$. Above $\omega_C$, the wire is animated of back-and-forth motion characteristic of the asynchronous regime and $\theta(t)$ displays oscillations. Because wires are superparamagnetic, the oscillation frequency is twice that of the excitation. The $\theta(t)$-transient behaviors aims to emphasize two important quantities here: the angle $\theta_B$ by which the wire returns after a period of increase (Fig. 3d) and the average angular velocity $\Omega = d\theta/dt$ (straight line in Fig. 3e). In the section IV dealing with biophysical applications, we will show that $\Omega(\omega,t)$ can be used to measure the viscosity of living media in transient and stationary regimes.

The simplest protocol to measure the fluid viscosity using the MRS technique is to determine experimentally the critical frequency $\omega_C$. Using wires of different lengths it is possible to verify the $L^{-2}$-dependence in Eq. 4, which also provides a strong support for the model validity. The *L*-variation method was shown to be satisfactory to describe the rheology of various fluids including viscous and viscoelastic materials.[4,23] For wires that are uniform in size, $\omega_C$ can be measured at increasing magnetic fields to reveal the $H^2$ dependence. Another method is to calculate the average angular velocity $\Omega(\omega)$ or the angle $\theta_B(\omega)$ from the $\theta(t)$-traces and to compare them with constitutive model predictions. For Newton and Maxwell fluids, it is found that $\Omega(\omega)$ expresses as:[4,21]

$$\omega \leq \omega_C \quad \Omega(\omega) = \omega$$
$$\omega \geq \omega_C \quad \Omega(\omega) = \omega - \sqrt{\omega^2 - \omega_C^2} \tag{5}$$

In Eq. 5, $\Omega(\omega)$ increases linearly with the frequency, passes through a maximum at $\omega_C$ before decreasing. The response of the probe over a broad spectral range appears as a resonance peak similar to that found in mechanical systems. The transition between the synchronous and asynchronous regimes is sometimes utilized for calibrating the wire susceptibility parameter $\Delta\chi$ in Eqs. 1 and 4.[4]

## 4. MICRO-RHEOLOGY OF INTRACELLULAR MEDIUM

With living cells, we address the question of the rheological properties of the cytoplasm, an issue that is relevant to understand how cells adapt to their environment and how forces are transmitted at the scale of individual cells.[26-28] Recent studies have suggested that the biomechanical response of the cytoplasm is analogous to that of a weak elastic solid.[29-32] Other work has led to the conclusion that the interior of living cells is best described as a purely viscous Newton fluid.[33,34] To test these hypotheses, cytoplasm viscosity measurements were performed in the passive and active modes on different cells lines including NIH/3T3 mouse fibroblsats, HeLa cervical cancer cells and A549 lung carcinoma epithelial cells.[35,36] These cell lines are representative of the cells used in the biomechanical studies carried out over the last decade.[27,28,33,37,38] From the active microrheology standpoint, the three cell lines were found to behave similarly when submitted to magnetic wire actuation. Here we review salient features obtained on the NIH/3T3 fibroblasts using the MRS technique and provide the first intracellular transient viscosity measurements.

We first establish the experimental conditions under which the wires enter in the cytoplasm. Two millions of fibroblast adherent cells at a 60% confluency were incubated with magnetic wires at a ratio 1:1, *i.e.* with an average of one wire per cell. Under such conditions, no perturbation of the cell morphology, cell cycle or viability was observed. Figs. 4 display an example of entry scenario inside the cytoplasm. We recall that the wires used here were not specifically modified to target the plasma membrane (they actually carry short neutral polymers at their outer surface).[35] Following incubation, the micron-sized wires need about 30 minutes to sediment and reach the adherent cell layer. Fig. 4b shows a 6 µm wire in contact with two neighboring fibroblast cells. At 60 min, the wire penetrates spontaneously in the upper cell and slowly crosses the plasma membrane in a process that lasts here around 25 minutes (Fig. 4d). In Fig. 4e and 4f, the wire is located in the cytoplasm and moves in phase with the cell motion. At the 1:1 cell-wire ratio, we found that not all the cells were loaded with wires, some cells showing several internalized objects and some none. Complementary immunofluorescence and TEM experiments have demonstrated that at 24 h the wires are not in membrane bound compartments, but instead disperse in the cytosol.[39]

Internalized wires of lengths 2 to 6 µm were studied as a function of the frequency using the MRS technique described in Section III.2. For $\omega$ between $10^{-3}$ and 10 rad s$^{-1}$, 3 to 5 minute movies were recorded and digitalized to retrieve the wire center-of-mass and orientation. At this time scale, the cells do not move much and it is assumed that the wires are probing the same intracellular volume throughout the experiment. In Figs 5a the orientation angle $\theta(t)$ of a 3.0 µm wire is plotted as a function of time at the frequency of 0.04 rad s$^{-1}$.

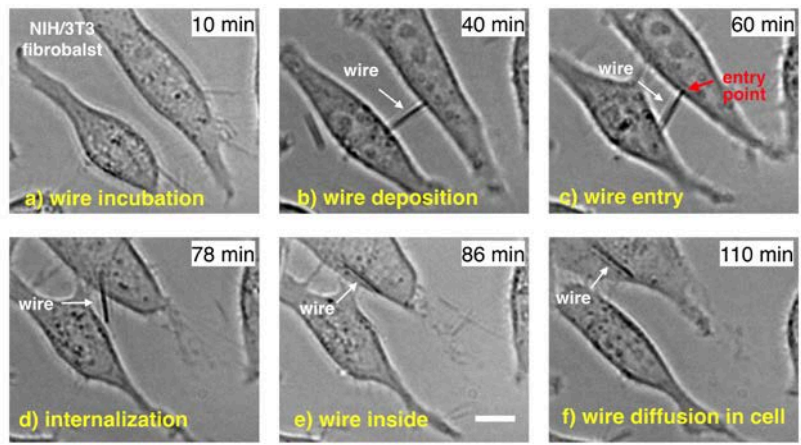

**Figure 4.** Snapshots of NIH/3T3 fibroblasts at different times after their incubation with magnetic wires. a) 10 min after the incubation, wires have not yet sedimented down to the cell layer. b) At 40 min a 6 µm wire in contact with two cells can be observed. c) The wire enters spontaneously inside the upper cell (rad arrow) 60 min after the incubation. d) The entry mechanism takes around 25 minutes. e and f) the wire is in the cell and diffuses freely in the cytoplasm. The bar is 5 µm.

There, $\theta(t)$ increases linearly, indicating that its motion is synchronous with the field. At higher frequency ($\omega$ = 0.41, 0.76 and 1.87 rad s$^{-1}$ in Figs. 5b, 5c and 5d respectively), the $\theta(t)$ traces display back-and-forth oscillations characteristic of the asynchronous regime. The critical frequency is here estimated at $\omega_C$ = 0.07 rad s$^{-1}$. Fig. 5e illustrates the movement of the wire in a single back-and-forth oscillation ($\omega$ = 0.68 rad s$^{-1}$). The images show that after a clockwise rotation in the 4 first panels, the wire comes back rapidly by 75 degrees in a counter-clockwise motion, indicating that the wire is actually hindered and cannot follow the magnetic field. On longer periods, the transient response in the asynchronous regime exhibits also irregular alternation of phases of constant average angular velocity $\Omega$ and phases of more rapid rotation. These features are characteristic for intermittency behaviors found in dynamic systems.[40] This succession of slowed down and accelerated rotations will be described in terms of time dependence viscosity.

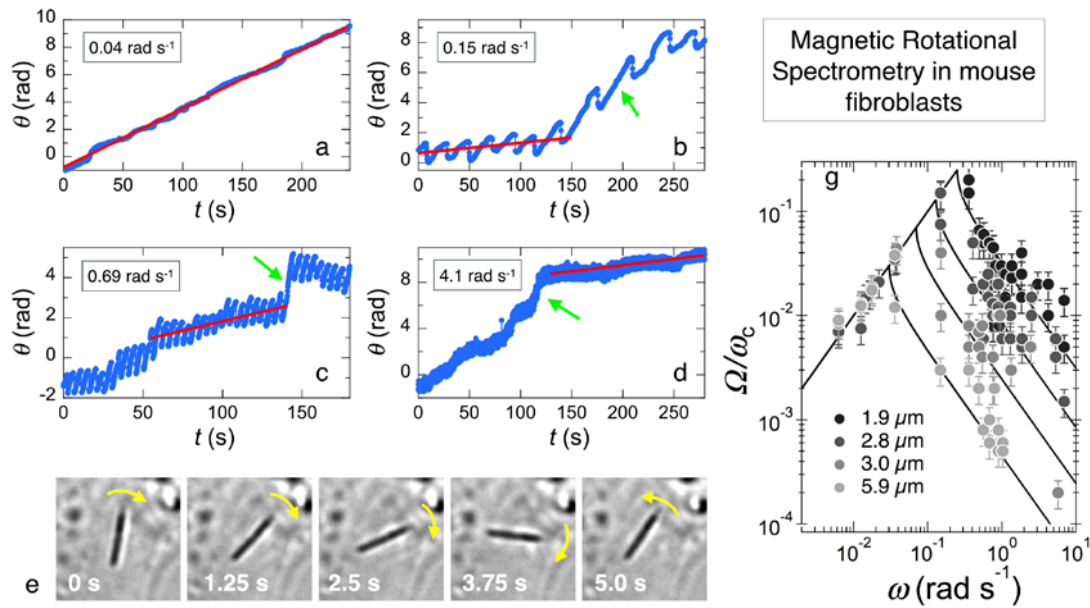

**Figure 5.** a-d) Time dependences of the wire orientation angle $\theta(t)$ for a 3 µm wire in the cytoplasm at frequencies $\omega$ = 0.04, 0.15, 0.69 and 4.1 rad s$^{-1}$. The straight lines in red display time regions where the average angular velocity is constant. e) Snapshots of a rotating wire in the asynchronous regime. The images show that after a clockwise rotation, the wire comes back in a counter-clockwise motion. g) $\Omega(\omega)$ measured for wires of various lengths together with the least-square adjustment using Eq. 5.

In a first approach, we focus on the long periods where the oscillations are regular and the average angular velocity constant, as illustrated by the red straight lines in the figures. Here, $\Omega(\omega)$ is defined as the average $\langle d\theta(t)/dt \rangle_t$. Fig. 5g displays these $\Omega(\omega)$-values for wires of different length. With increasing frequency, the average velocity increases linearly, passes through a maximum at $\omega_C$ and then decreases. The transition corresponds to the change of rotation regime from synchronous rotation to back-and-forth oscillations, as mentioned previously. Note also that as expected from Eq. 4 $\omega_C$ increases as the length of the wire decreases. The data in Figs. 5g were successfully adjusted using Eq. 5 and in agreement with the constitutive model predictions for viscous and viscoelastic liquids. In the selected examples, $\omega_C$ varies from 0.03 to 0.2 rad s$^{-1}$ and is associated to static viscosities between 20 and 80 Pa s. These results also rule out the hypothesis of the intracellular medium being a weak elastic gel, as suggested recently.[6,29,30] In this context we have shown that wires embedded in elastic gels or equivalently in soft solids behave differently and that their average rotation velocities $\Omega(\omega) = 0$, which is not the case for cells.[6] An analysis of the oscillation amplitudes $\theta_B(\omega)$ from Fig. 5b-d reveals that the cytoplasmic rheological properties are viscoelastic and characterized by an elastic modulus of the order of 10 Pa for the fibroblasts.[35]

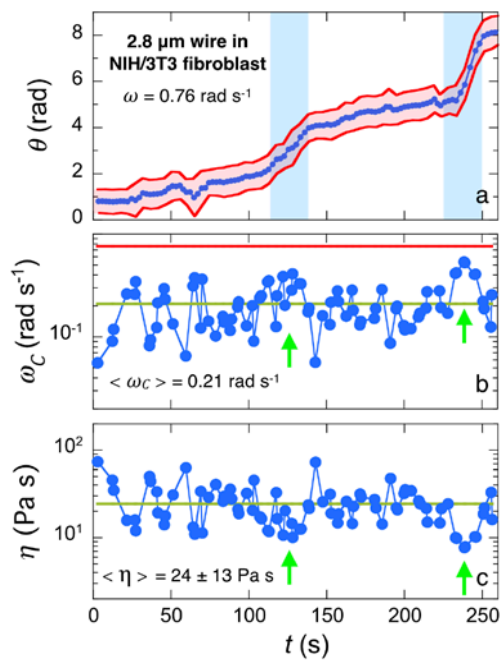

**Figure 6.** a) Average angle $\theta(t)$ of a 2.8 μm wire located in the cytoplasm of a NIH/3T3 fibroblast and submitted to a 10 mT field at $\omega$ = 0.76 rad s$^{-1}$. The envelope in red delimits the back-and-forth oscillations characteristic of the asynchronous rotation. b) Instantaneous critical frequency $\omega_C(t)$ obtained from Eq. 5; the horizontal red and green lines denote values of the applied and average critical frequencies, respectively. c) Instantaneous viscosity $\eta(t)$ derived from Eq. 4 showing strong fluctuations as a function of the time. The arrows in green point time regions where the local critical frequency is increased and the local viscosity is decreased.

It should be noticed that at frequency $\omega > \omega_C$ in Fig. 5g the data exhibit some scattering that could come from time-dependent effects and be related to the cellular activity. To get more insight into this effect, the $\theta(t)$-traces have been analyzed by determining the instantaneous average angular velocity $d\theta(t)/dt$ at the time scale of each oscillation. Fig. 6a shows the average angle $\theta(t)$ obtained for a 2.8 μm wire inserted in the cytoplasm of a fibroblast in the asynchronous regime ($\omega$ = 0.76 rad s$^{-1}$). The envelope of the back-and-forth oscillations is also represented and delimits the zone of the accessible wire rotations in this assay. In the diagram, we also note that around 125 s and 240 s (shown in blue), the wire accelerates and the average rotation frequency increases. It is then assumed that Eq. 5 first defined to describe the stationary regime is also valid at the scale of the oscillations, leading an expression of the form $d\theta(t)/dt = \omega - \sqrt{\omega^2 - \omega_C(t)^2}$. In the previous expression, the frequency of the rotating field is fixed and the fluctuations in the wire response are solely due the viscosity change of the medium surrounding the wire. With these assumptions, the instantaneous critical frequency $\omega_C(t)$ and viscosity $\eta(t)$ can be estimated and plotted *versus* time (Fig. 6b and 6c respectively). To get the viscosity from the critical frequencies, Eq. 4 is used. The results that emerge from this approach are first that the cytoplasm viscosity of a living cell exhibits strong temporal fluctuations associated to the normal cellular activity. In this particular assay the standard deviation (13.6 Pa s) represents more than 50% of the average viscosity. Second it is interesting to note that in the light blue area where the wire accelerates, the critical frequency is higher that the average and the viscosity is lowered by 50 to 70 % with respect to the average. Additional time-resolved microrheology experiments performed on HeLa cervical cancer cells and A549 lung carcinoma epithelial cells have confirmed these transient patterns. In conclusion, we have found that using reasonable assumptions, the magnetic rotational spectrometry can provide insight in the time biomechanical response of living cells, showing in particular strong fluctuations of the intracellular viscosity.

# 5. CONCLUSION

We have shown here that micron-sized wires with superparamagnetic properties are remarkable probes to determine quantitatively rheological properties of fluid and gel-like materials. Comparing micro- and macrorheology, we point out that under certain magnetic field conditions, the wires can be compared to cone-and-plate or Couette shear devices mounted on a stress-controlled rheometer. One key advantage of these magnetic wires is that they can be monitored at frequencies as low as 1 mHz, allowing to test materials with long relaxation times and reach the liquid-like behavior of most viscoelastic liquids. The use of superparamagnetic wires coupled with the magnetic rotational spectroscopy technique is able to capture the rheological nature of the intracellular medium of living cells, which is that of a viscoelastic liquid. In conclusion, wire-based microrheology is a powerful technique able to determine the rheological nature of viscoelastic materials, and provide quantitative rheological parameters such as the static viscosity or the shear elastic modulus.

## ACKNOWLEDGMENTS

A. Cebers, L. Chevry, F. Mousseau, E. Oikonomou, L. Vitorazi are acknowledged for fruitful discussions. Interns who participated to the research, C. Leverrier, A. Conte-Daban, C. Lixi, L. Carvhalo, M. Najm and R. Chan are also acknowledged. This research was supported in part by the Agence Nationale de la Recherche under the contract ANR-13-BS08-0015 (PANORAMA), ANR-12-CHEX-0011 (PULMONANO) and ANR-15-CE18-0024-01 (ICONS, Innovative polymer coated cerium oxide for stroke treatment) and by Solvay® Singapore. ANR and CGI are also acknowledged for their financial support of this work through Labex SEAM (Science and Engineering for Advanced Materials and devices) ANR 11 LABX 086, ANR 11 IDEX 05 02.